\documentclass[12pt,linenumbers]{article}

\usepackage[a4paper,left=30mm,right=20mm,top=20mm,bottom=20mm]{geometry}
\usepackage{graphics}
\usepackage{amsmath}
\usepackage{epsfig}
\usepackage{cite}
\usepackage{lineno}
\usepackage[varg]{txfonts}
\newcommand{\be}{\begin{eqnarray}}
\newcommand{\ee}{\end{eqnarray}}
\title{Mid-rapidity dependence of hadron production in $p-p$ and $A-A$ collisions.
}

\author{A.I. Malakhov, G.I. Lykasov} 
\begin{document}
\date{}
\maketitle

\begin{center}

{Joint Institute for Nuclear Research, Dubna 141980, Moscow region, Russia

\vspace{0.5cm}

%artemenkov.denis@gmail.com\\
lykasov@jinr.ru\\
malakhov@lhe.jinr.ru\\
%jerus@jinr.ru
}

\end{center}

\vspace{0.5cm}

\begin{center}

{\bf Abstract }

\end{center} 

%\indent
%{\bf
The calculation of inclusive spectra of pions produced in $pp$ and $AA$ collisions as a function of rapidity $y$
is presented within the self-similarity approach. 
It is shown that at not large rapidities $y$ one can obtain the analytical form of the self-similarity
function $\Pi(y,p_t)$ dependent of $y$ and hadron transverse momentum $p_t$. A satisfactory description of data on the
rapidity spectra at $|y|\leq$ 0.3 is illustrated within a good agreement. 
The universal energy dependence of these spectra is also shown.

\vspace{0.5cm}

\noindent
%PACS number(s): 12.15.Ji, 12.38.Bx, 13.85.Qk

%\newpage
\indent
%%%%%%%%%%%%%%%%%%%%%%%%%%%%%%%%%%%%%%%%%%%%%%%%%%%%%%%%%%%%%%%%%%%%%%%%%%%%%%%%%%%
%%%%%%%%%%%%%%%%%%%

\section{Introduction}
\label{intro}

The approach based on similarity of inclusive spectra of particles produced in hadron-hadron collisions suggested in pioneering papers
\cite{Fermi:1950,Pomeran:1951,Landau:1953,Hagedorn:1965} was developed in \cite{c3,c4,4,5}. In \cite{c4,5} the similarity of these spectra 
as a function of similarity parameter $\Pi$ dependent of the initial energy $\sqrt{s}$  in the c.m.s of colliding particles and transverse 
mass $m_{ht}$ of produced hadrons at zero rapidity $y=$0 was demonstrated. A simple form of inclusive spectra was used in \cite{c4,4,5} 
to describe satisfactorily the spectra at low values of $m_{ht}$. Further development of this approach was presented 
in our papers \cite{1,2,3}, where the description of $m_{ht}$-spectra was extended at larger values of transverse momenta and initial energies 
up to a few TeV including both contributions of quarks and gluons to these spectra. It has been shown \cite{1} the direct relation of $\Pi$ to the
Mandelstam variables $s,t$ and its non factorized form as a common function of $s$ and $m_{ht}$, which is very significant at not large 
initial energies $\sqrt{s}<$ 10 GeV and becomes factorized at larger $\sqrt{s}$. In fact, this is an advantage of the approach based on kinematics
of four-momentum velocities considered in \cite{c3,c4,4,5}, where the parameter $\Pi$ was obtained using the conservation law of four-momenta
and quantum numbers of initial and produced particles, and the minimization principle. At zero rapidity $y$=0 the form for $\Pi$ was obtained 
analytically \cite{5}. 

Let us note, that at non zero rapidity there are many theoretical models describing the inclusive spectrum of hadrons produced in $A-A$ collisions
as a function of $y$ and $m_{ht}$, see for example, \cite{Thermmod:1993,Wilk:2000,Bugaev1:2002,Bugaev2:2002,Cleymans:2013} and references there in. 
There are also experimental data
on these distributions and their fits \cite{7,10,11}. However, by modeling or fitting the rapidity and transverse momentum dependence there is no
their universal energy dependencies in these papers. 

In this paper we extend approach offered in \cite{1} at the non zero rapidity region and calculate analytically the similarity 
parameter $\Pi$ as a function of $y$ and $m_{ht}$ without any additional parameters. 
Then,  we calculate the $y$-dependence of inclusive spectra of pions produced in $pp$ and $AA$ collisions and describe satisfactorily data 
at $|y|\leq$ 0.3 in a wide region of initial energies. We also have confirmed that the distributions over $y$ and $m_{ht}$ have the universal 
energy dependence at low values of these variables, as it has already been shown in \cite{1}.

\section{The parameter or function of self-similarity $\Pi$.}
\label{sec:1}

The inclusive production of hadron 1 in the interaction of nucleus A with nucleus B:     
\begin{equation}
A  +   B  \rightarrow  1  + \ldots ,                                           
\label{eq:n1}                                                                
\end{equation}                                              
is satisfied by the conservation law of four-momentum in the following form:
\begin{equation}
{(N_AP_B + N_BP_{B} - p_1)}^2 = 
{(N_Am_0 + N_B m_0 + M)}^2 ,
\label{eq:n2}
\end{equation}
where $N_A$ and $N_B$ are the fractions of four momenta transmitted by the nucleus A and nucleus B; $P_A , P_B , p_1$ are four momenta of the nuclei 
A and B and particle 1, respectively; $m_0$ is the mass of the nucleon; M is the mass of the particle providing the conservation of the baryon 
number, strangeness, and other quantum numbers.
For $\pi$-mesons $m_1 = m_\pi$  and M = 0.
For antinuclei and $K^-$-mesons  $M = m_1$.
For nuclear fragments $M = - m_1$.
For $K^+$-mesons $m_1 = m_K$ and $M = m_\Lambda  - m_K$, $m_\Lambda$ is the mass of the $\Lambda$-baryon.

     In \cite{4} the parameter of self-similarity is introduced, which allows one to describe the differential cross section of the yield of a large class of 
particles in relativistic nuclear collisions:
\begin{equation} 
\Pi=\min[\frac{1}{2} [(u_A N_A + u_B N_B)^2]^{1/2} ,
\label{eq:n3} 
\end{equation}                                            
where $u_A$ and $u_B$ are four velocities of the nuclei A and B. 

     Then, the inclusive spectrum of the produced particle 1 in AA collision can be presented as the general universal function dependent of the 
self-similarity parameter: 
\begin{eqnarray}
E d^3 \sigma/dp^3~=~A_A^{\alpha(N_A} \cdot A_B^{\alpha(N_{B}} \cdot F(\Pi)
%\exp(-\Pi/C_2),
\label{eq:n4} 
\end{eqnarray}
where $\alpha(N_A)=1/3 + N_A/3$, $\alpha(N_B)=1/3 + N_B/3$, 
%$C_1=1.9 \cdot 10^4 $ mb $\cdot$ GeV$^{-2}$ $\cdot$ c$^3$ $\cdot$ st$^{-1}$ 
and $F$ is the function, its form is presented in \cite{1}:
%$C_2 = 0.125 \pm 0.002$.
\begin{eqnarray}
F(\Pi)~=~[A_q exp(-\Pi/C_q)~+~A_g\sqrt{m_{1t}} exp(-\Pi/C_g)
%\\
%\nonumber
\sigma_{nd}/g((s/s_0)^{\Delta})]\cdot g(s/s_0)^{\Delta}~.
\label{eq:F} 
\end{eqnarray}
Here $\Delta=\alpha_P(0)-1\simeq 0.12$ is the excess of the sub critical Pomeron intercept
over 1; $g=21 mb$ - constant, which was calculated within the "quasi-eikonal" approximation
\cite{c11}. The constants $A_q=$3.68~(GeV$/$c)$^{-2}$, C$_q=$0.147; $A_g=$1.7249~(GeV$/$c)$^{-2}$, $C_g$=0.289
were obtained in \cite{c13,ALM:2015}.

\section{Analytical solution for self-similarity parameter}

     An analytical solution for the self-similarity parameter П was found in \cite{5}. Here we give a more detailed derivation of the parameter and consider 
its behavior at small values of $y<<$ 1.
     Equation (2) can be written as follows:
\begin{equation} 
N_A\cdot N_B - \Phi_A\cdot N_A - \Phi_B\cdot N_B = \Phi_M ,
\label{eq:n5} 
\end{equation}

where relativistic invariant dimensionless values have been introduced:
$$ \Phi_A = [(m_1/m_0) \cdot (u_Au_1) + M/m_0]/[(u_Au_B)- 1]$$
$$ \Phi_{B} = [(m_1/m_0) \cdot (u_Bu_1) + M/m_0]/[(u_Au_B)- 1]$$
$$ \Phi_M = (M^2 - m^2_1)/[2m_0^2((u_Au_B)- 1)].$$

It was shown that at  $y = 0$
\begin{eqnarray}
\Phi_A =  \Phi_B = \Phi, \\
\nonumber
N_A=N_B = N = {1 + [1 + (\Phi_M/ \Phi^2)]^{1/2} }·\Phi. \\
\nonumber
\Pi = N\cdot ch(Y).
\label{eq:n6} 
\end{eqnarray} 
The scalar product of four-dimensional velocities is related to the rapidity of initial particles $Y$ and 
the rapidity $y$ of the produced hadron $1$:
\begin{eqnarray}
(u_Au_B) = ch(2Y), \\
\nonumber
(u_A\cdot u_1)~=~(m_{1t}/m_1)\cdot ch(-Y-y) = (m_{1t}/m_1)\cdot ch(Y+y), \\
\nonumber
(u_B\cdot u_1)~=~(m_{1t}/m_1)\cdot ch(Y-y), 
\label{eq:n6}
\end{eqnarray}
%where
%$$
%ch(Y+y)  =ch(Y)∙ch(y) + sh(Y)∙sh(y)
%$$
Here $m_{1t} = (m_1^2 + p_{1t}^2)^{1/2}$ is the transverse mass of the particle 1.
%%%%%%%%%%%%%%%%%%%%%%%%%%%%%%%%%%%%%%%%%%%%%%%%%%%%%%%%%%%%%%%%%%%%%%%%%%%%%%%%%%%%%%%%%%%%%
If $y<<$ 1, then one can neglect $sh(y)$ compared to $ch(y)$ and approximately we get the following:
\begin{eqnarray}
 (u_A\cdot u_1)\simeq (u_Bu_1)\simeq (m_{1t}/m_1)\cdot ch(y)ch(Y),
\label{eq:n6}
\end{eqnarray}
And in this case:
\begin{eqnarray}
\Phi~=~\Phi_A~=~\Phi_B~=~[(m_1/m_0)∙(u_Au_1) + M/m_0]/[(u_Au_B)-1]\simeq\\
\nonumber
 [(m_1/m_0)\cdot (m_{1t}/m_1)\cdot ch(y)\cdot ch(Y)+ M/m_0]/[ch(2Y)-1] = \\ 
\nonumber
   = {(1/m_0)[m_{1t}\cdot ch(y)\cdot ch(Y) + M]}\cdot [1/(2sh^2(Y))]. \\
\nonumber
 \Phi_M = (M^2 - m_1^2)/(4m_0^2sh^2(Y)).
\label{eq:n7}
\end{eqnarray}
Thus at $y<<$ 1
\begin{eqnarray}
N = {1 + [1 + ( \Phi_M/\Phi^2)]^{1/2} } \Phi,
\label{eq:n7}
\end{eqnarray}
where 
\begin{eqnarray}
\Phi\simeq {(1/m_0)[m_{1t}\cdot chy\cdot chY + M]}∙[1/(2sh^2(Y))],\\
\nonumber
\Phi_M = (M^2 - m_1^2)/(4m_0^2sh^2(Y)).
\label{eq:n8}
\end{eqnarray}

     Since this equation $(u_Au_1)$ does not depend on $m_A$, it is valid for any hadrons and nuclei:
\begin{eqnarray}
(u_A\cdot u_1) = (P_A/m_A)(P_1/m_1) = (E_A\cdot E_1/m_A\cdot m_1) - (p_A\cdot p_1/m_A\cdot m_1) \\
\nonumber
 = m_A∙ch(Y)∙m_{1t}∙ch(y)/(m_A∙m_1) + m_A∙sh(Y)∙m_{1t}∙sh(y)/(m_A∙m_1) = \\
\nonumber 
= (m_{1t}/m_1)\cdot (ch(Y)\cdot ch(y)+sh(Y)\cdot sh(y)) = (m_{1t}/m_1)\cdot ch(Y+y).
\label{eq:n9}
\end{eqnarray}
     Therefore, we conclude that our approach is also valid for projectile π mesons.

\section{Rapidity distribution of pions at low $y$.}

Using the relation of rapidities $Y$ and $y$ of initial particles and produced hadron $1$ respectively 
to the Mandelstam variables $s$ and $t$ \cite{1}, one can get the following form of the similarity parameter 
$\Pi$ for small but non zero rapidity $y$:
\begin{eqnarray}
\Pi(s,m_{1t},y)\simeq\frac{m_{1t}ch(y)}{2m_0(1-4m_0^2/s)}
\left\{1+\sqrt{1+\frac{M^2-m_1^2}{m_{1t}^2ch^2(y)}(1-4m_0^2/s)}\right\}~.
\label{eq:n10}
\end{eqnarray}
For $pp\rightarrow h + X$ inclusive processes the relativistic invariant differential cross at small but non zero rapidity $y$ will
have the following form:
\begin{eqnarray}
E_h\frac{d^3 \sigma_{NN}}{d^3p_h}~\equiv \frac{1}{\pi}\frac{d\sigma}{dm_{1t}^2dy}=F(\Pi(s,m_{1t},y)),
\label{eq:n11}
\end{eqnarray}
where $F(\Pi(s,m_{1t},y))$ is given by Eq.~\ref{eq:F} but with $\Pi(s,m_{1t},y))$ determined by Eq.~\ref{eq:n10}. 
For $AA\rightarrow h + X$ processes the differential cross section is presented by Eq.~\ref{eq:n4}.
The integral of Eq.~\ref{eq:n11} or Eq.~\ref{eq:n4} over the transverse mass of the produced hadron $m_{1t}$ results in
the rapidity $y$ dependence of the cross section of hadrons produced in $pp$ or $AA$ collision, respectively. 
Finally the rapidity distribution can be presented in the following form:
\begin{eqnarray}
\frac{d\sigma_{NN}}{dy}~=~2\pi \int F(\Pi(s,m_{1t},y)) m_{1t} dm_{1t}
\label{eq:n12}
\end{eqnarray}
\begin{figure}[hbtp] 
\begin{center}
\begin{tabular}{ccc}
%\resizebox{0.55\textwidth}{!}{%
\includegraphics[width=0.277\textwidth]{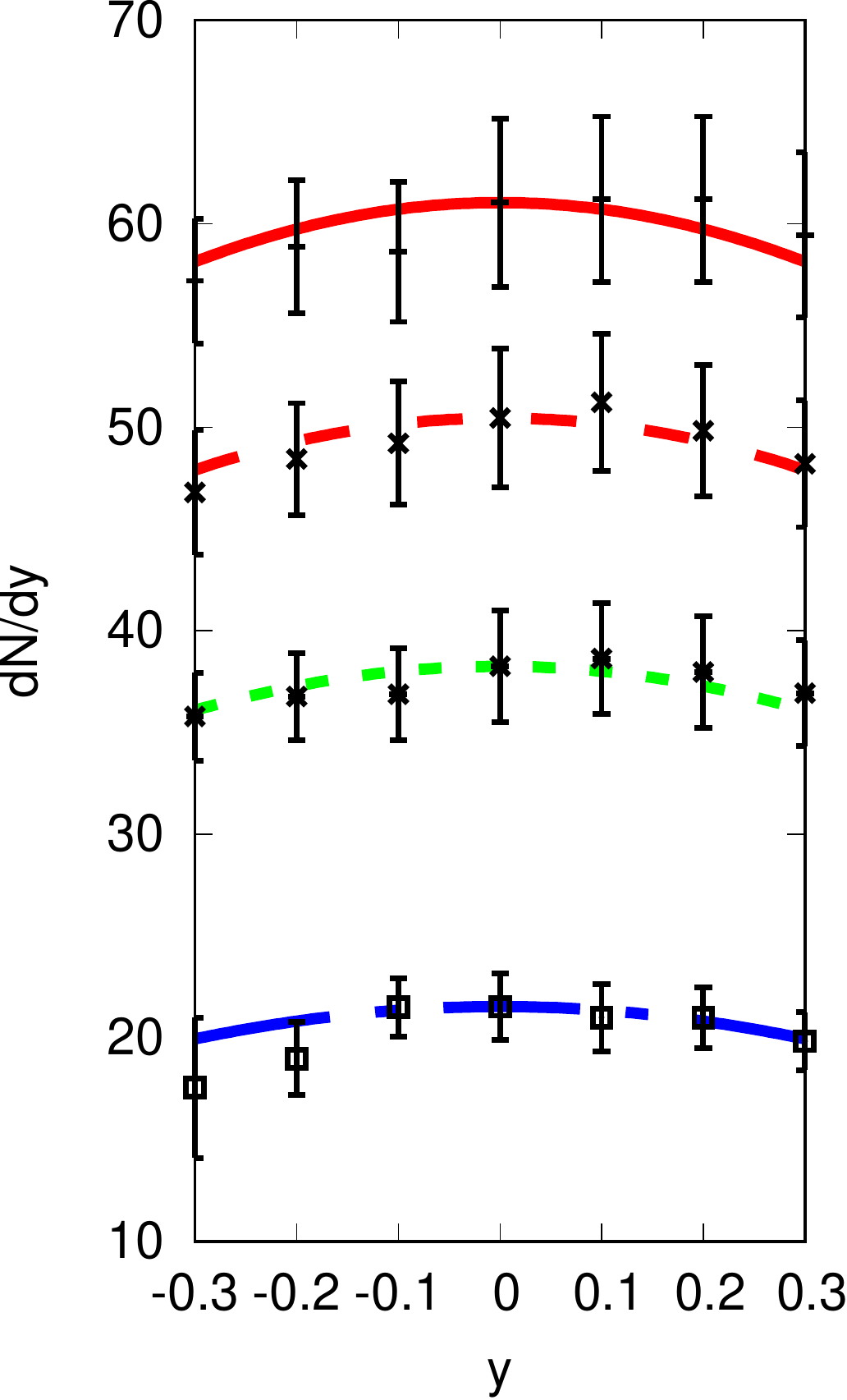}&
%sp_y_AGS.pdf}
%\resizebox{0.55\textwidth}{!}{%
\includegraphics[width=0.3\textwidth]{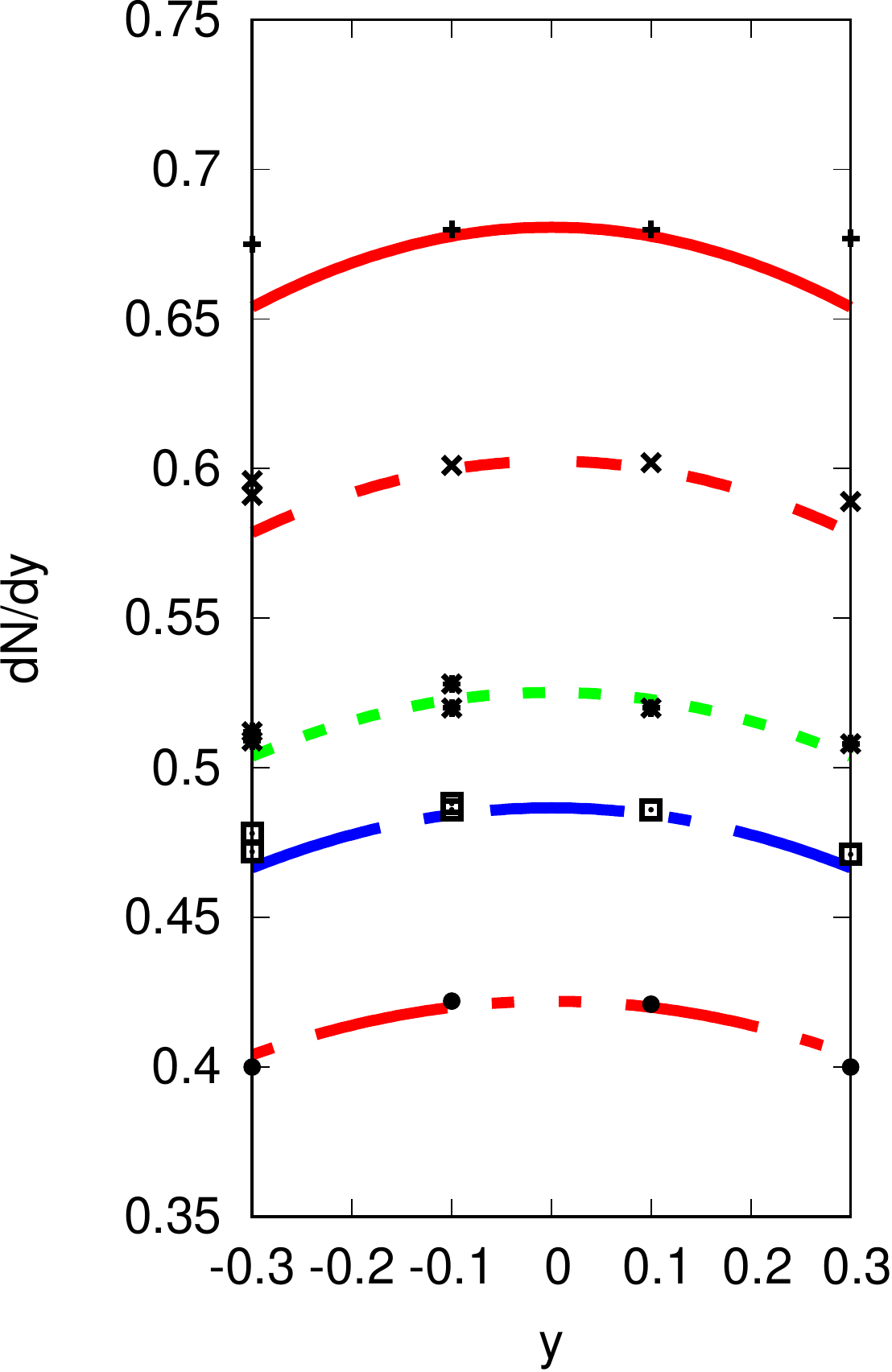}&
%sp_y_NA61.pdf}
%\resizebox{0.55\textwidth}{!}{%
\includegraphics[width=0.292\textwidth]{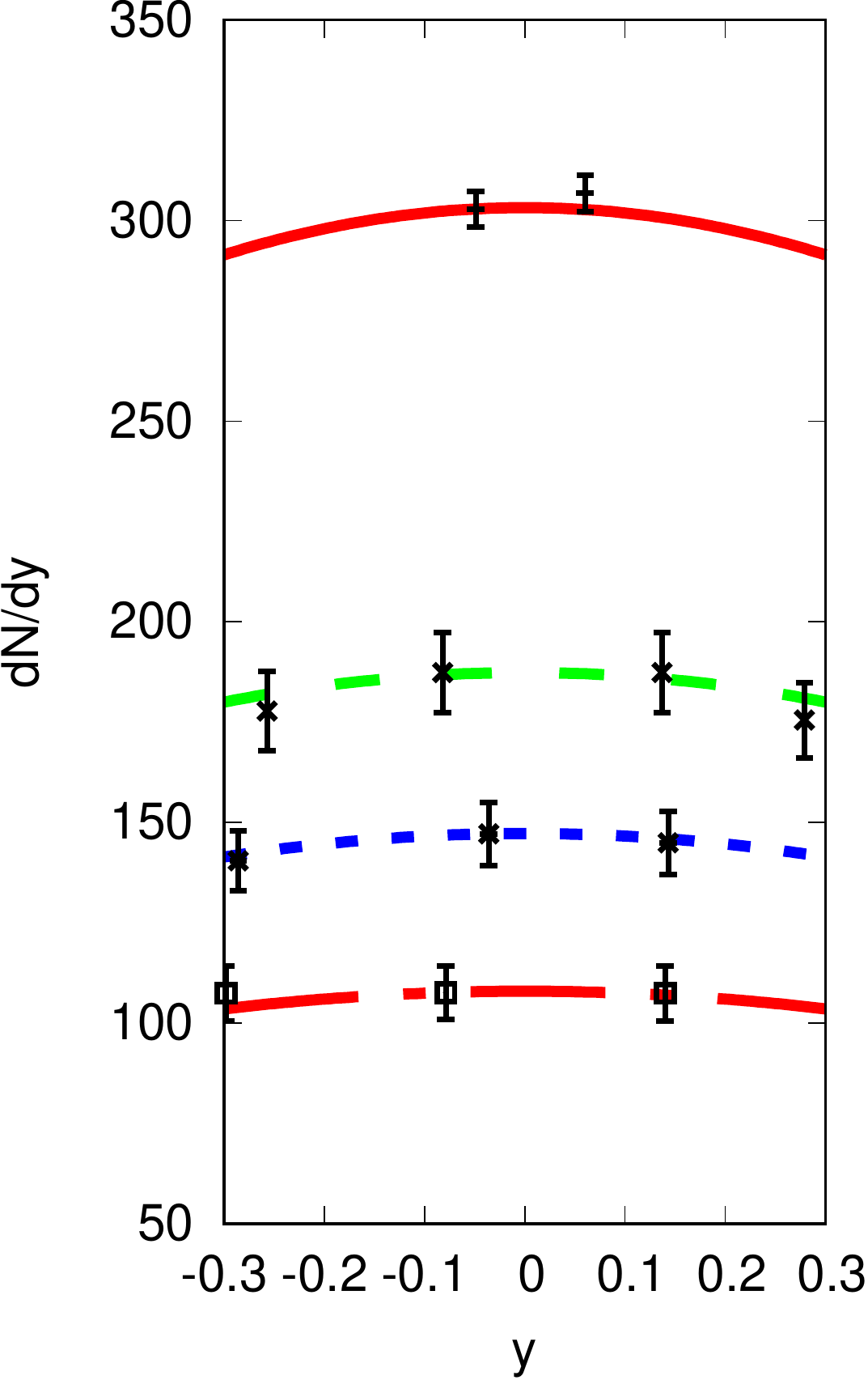}
%}
\end{tabular}
\end{center}
 \caption{
Left: pion rapidity $y$-spectra in $AuAu$ collision at $\sqrt{s}=$ 4.31 A GeV (solid line), 3.84 A GeV (long dash line),
 3.32 A GeV (shirt dash line), 2.7 A GeV (dashed-dotted line)
or the initial kinetic energies per nucleon about $E_{kin}=$ 8, 6, 4, 2 GeV, respectively. They are compared to the AGS data \cite{7}.
Middle: pion $y$-spectra in $AuAu$ collision (solid line, RHIC data) at $\sqrt{s}=$ 200 A GeV 
and $PbPb$ collision (SPS data) at $\sqrt{s}=$ 17.2 A GeV (long dashed line), 12.3 A GeV (short dashed line) and 8.7 A GeV (dashed-dotted line) 
The RHIC and SPS data were taken from \cite{11}. 
Right: pion $y$-spectra in $pp$ collisions at the initial momentum 
$P_{in}=$158 GeV$/$c (solid line), 80  GeV$/$c (long dashed line), 40 GeV$/$c (short dashed line), 31  GeV$/$c (dashed-dotted line), 20 GeV$/$c
(dashed-double dotted line) compared to the NA61/SHINE  data \cite{10}.  
}  
\label{fig1}
\end{figure}
In Fig.~(\ref{fig1}) the $y$-distributions of pions produced in $pp,AuAu$ and $PbPb$ collisions are presented
in the wide region of initial energies.
A satisfactory description of data with a precision less than 10\% is shown in the rapidity range of the produced particles
$|y|<$0.3.

Note that the main contribution to $d\sigma/dy$ given by Eq.~\ref{eq:n12} comes from the first term of Eq.~\ref{eq:F} at low values of $y$,
as our calculations have shown. Therefore, the rapidity distribution can be presented in the following approximated form :
\begin{eqnarray}
\frac{d\sigma_{NN}}{dy}~\simeq~A_A^{\alpha(N_A)} \cdot A_B^{\alpha(N_(B)}A_q g(s/s_0)^{\Delta}m_1 
%\\
%\nonumber
[\frac{m_1}{ch(y)}+\frac{2m_0\delta}{ch^2(y)}]C_q\exp(-m_1ch(y)/(2m_0\delta C_q))~,
\label{eq:ydistr}
\end{eqnarray}
where $\delta=1-4m^2_0/s$. The Eq.~\ref{eq:ydistr} is similar to $d\sigma_{NN}/dy$ obtained in \cite{Thermmod:1993} within the
thermal model including the longitudinal and transverse flow. The difference between our rapidity distribution and the one 
considered in \cite{Thermmod:1993} is the following. We do not include the nuclear thermal effects, which can change the $y$-dependence, mainly 
at $|y|>$ 0.3. Our approach can be applied at $|y|\leq$ 0.3 rather satisfactorily, as it is shown from Fig.~(\ref{fig1}).
Eq.~\ref{eq:ydistr} results in the universal energy dependence of $d\sigma/dy$, as $(s/s_0)^{\Delta}$.
More complicated energy dependence of $d\sigma/dm_{1t}$ was obtained in our previous paper \cite{1}.  

\section{Conclusion}

In this paper we have extended the self-similarity approach of analysis of hadron production in $pp, pA$ and $AA$ collisions,
which firstly has been suggested in \cite{c3,4,5} and developed in \cite{1} strictly at zero rapidity $y=0$ of produced
hadrons, to the non zero rapidity region. This extension was obtained analytically using the conservation law of four-momenta and quantum 
numbers of initial and final particles. The validity of our results concerns the rapidity interval $|y|<$ 0.3, as it was shown by satisfactory description 
of the data on the pion production in $pp$ and $AA$ collisions ($\leq $) 0.3 within the wide region of initial energies.
Moreover, we have got the universal energy dependence of $y$-spectra using the excess of the sub critical Pomeron intercept
over 1, which is known very well from the satisfactory description of many data on the hadron production in $pp$ collision.

{\bf Acknowledgements.}

\begin{sloppypar} 
We are very grateful to K.A. Bugaev, M. Gumberidze, M. Gadzicky, R. Holzmann, G. Kornakov, A. Rustamov
 for extremely helpful discussions.   
\end{sloppypar}

%%%%%%%%%%%%%%%%%%%%%%%%%%%%%%%%%%%%%%%%%%%%%%%%%%%%%%%%%%%%%%%%%%%%%%%%%%%%%%%%%%%%%%%%%%%%%

\end{document}